\documentstyle[epsfig,psfig]{mn}

\def\ref{\par\noindent\hang}

\def\etal{{et al.\thinspace}}
\def\cf{{\em cf.\ }}
\def\eg{{\em e.g.\ }}

\def\spose#1{\hbox to 0pt{#1\hss}}
\def\approxlt{\mathrel{\spose{\lower 3pt\hbox{$\sim$}}
        \raise 2.0pt\hbox{$$<$$}}}
\def\approxgt{\mathrel{\spose{\lower 3pt\hbox{$\sim$}}
        \raise 2.0pt\hbox{$>$}}}

\def\multleft#1{\hbox to size{\vbox {\halign {\lft{##}\cr #1}}\hfill}\par}
\def\multright#1{\hbox to size{\vbox {\halign {\rt{##}\cr #1}}\hfill}\par}

\def\today{\ifcase\month\or January\or February\or March\or April\or May\or
      June\or July\or August\or September\or October\or November\or December\fi
      \space\number\day, \number\year}
\def\$<${\thinspace}
\def\s{\hbox{\phantom{5}}}      

\def\boxit#1{\vbox{\hrule\hbox{\vrule\kern3pt\vbox{\kern3pt
          #1 \kern3pt}\kern3pt\vrule}\hrule}}

\def\ga{{\rm\thinspace gauss}}
\def\K{{\rm\thinspace K}}

\def\km{{\rm\thinspace km}}
\def\kg{{\rm\thinspace kg}}

\def\Lsun{\hbox{$\rm\thinspace L_{\odot}$}}

\def\m{{\rm\thinspace m}}
\def\Mpc{{\rm\thinspace Mpc}}
\def\Msun{\hbox{$\rm\thinspace M_{\odot}$}}

\def\s{{\rm\thinspace s}}
\def\yr{{\rm\thinspace yr}}

\def\Hz{{\rm\thinspace Hz}}

\def\Msunpyr{\hbox{$\Msun\yr^{-1}\,$}}

\def\mic{{\rm\thinspace $\mu$m}}
\def\mJy{{\rm\thinspace mJy}}

\def\la{\mathrel{\mathchoice {\vcenter{\offinterlineskip\halign{\hfil
$\displaystyle##$\hfil\cr<\cr\sim\cr}}}
{\vcenter{\offinterlineskip\halign{\hfil$\textstyle##$\hfil\cr
<\cr\sim\cr}}}
{\vcenter{\offinterlineskip\halign{\hfil$\scriptstyle##$\hfil\cr
<\cr\sim\cr}}}
{\vcenter{\offinterlineskip\halign{\hfil$\scriptscriptstyle##$\hfil\cr
<\cr\sim\cr}}}}}

\def\ga{\mathrel{\mathchoice {\vcenter{\offinterlineskip\halign{\hfil
$\displaystyle##$\hfil\cr>\cr\sim\cr}}}
{\vcenter{\offinterlineskip\halign{\hfil$\textstyle##$\hfil\cr
>\cr\sim\cr}}}
{\vcenter{\offinterlineskip\halign{\hfil$\scriptstyle##$\hfil\cr
>\cr\sim\cr}}}
{\vcenter{\offinterlineskip\halign{\hfil$\scriptscriptstyle##$\hfil\cr
>\cr\sim\cr}}}}}

\begin{document}
\hsize=6truein

\title[Sub-mm observations of high-z quasars]
{Submillimetre observations of luminous $\boldmath z > 4 $ radio-quiet quasars 
and the contribution of AGN to the submm source population
}
\author[R.G. McMahon \etal]
{\parbox[]{6.5in}
{Richard G. McMahon$^1$, Robert S. Priddey$^1$, 
Alain Omont$^2$,
Ignas Snellen$^1$, 
Stafford Withington$^3$ 
}\\
1. \it Institute of Astronomy, Madingley Road, Cambridge CB3 0HA, UK\\
2. Institut d'Astrophysique de Paris, CNRS, 98 bis Bd. Arago, F-75014, Paris, France\\ 
3. Mullard Radio Astronomy Observatory, Cavendish Laboratory, Cambridge CB3 0HE, UK\\
\rm email: rgm, rpriddey@ast.cam.ac.uk, omont@iap.fr, snellen@ast.cam.ac.uk, stafford@mrao.phys.ast.cam.ac.uk}

\date{Accepted 1999 June 29}

\maketitle

\begin{abstract}
  We present sensitive 850\mic\ SCUBA photometry of a statistically-complete 
  sample of six of the most luminous ($M_{\rm{B}} < -27.5$; 
  $\rm \nu L_{\nu}{\rm (1450\thinspace\AA)} \ga 10^{13}\Lsun$), 
  $z>4$ radio-quiet
  quasars, reaching noise levels ($1\sigma \sim 1.5\mJy$) comparable
  with the deep blank sky surveys. These observations probe the rest
  frame far infrared region ($\sim$150\mic), at luminosity
  levels for thermal sources comparable with those that {\it IRAS} permitted
  for low redshift quasars.  
  One quasar (BR2237$-$0607; $z=4.55$) is
  detected at 850\mic\ with a flux of 5.0$\pm$1.1mJy ($4.5\sigma$),
  whilst a second (BR0019$-$1522; z=4.52) has a detection
  at the $2\sigma$ level. 
  When combined with our previous millimetre
  measurements of z$>$4 quasars, we find that there is a large
  range (5--10) in far infrared (FIR) luminosity ($L_{\rm FIR}$) at fixed UV
  luminosity, and that the typical quasar has a $L_{\rm FIR}$ and
  mass of cool (50\K) dust similar to that of the archetyepal low
  redshift (z=0.018) ultraluminous {\it IRAS} galaxy(ULIRG)
  Arp220 ($L_{\rm FIR}\sim5\times10^{12}\Lsun$; 
  $M_{\rm d}{\rm(cool)}\sim10^{8}\Msun$). 
  If one assumes a fiducial FIR luminosity of $5\times10^{12}\Lsun$ for 
  for all quasars with $M_B<-23$,
  we find that $\ga$15 per cent of
  the sources in the SCUBA deep surveys could be classical broad-lined
  radio-quiet AGN. Thus if one considers the observed ratio of Seyfert II to
  Seyfert I galaxies at low redshift 
  and any contribution from totally
  optically obscured AGN, a significant fraction of the SCUBA source 
  population will harbour AGN and hence the inferred star formation
  rates from submm fluxes may be overestimated
  if the active nuclei are bolometrically dominant or the IMF is top
  heavy.

\end{abstract}


\section{INTRODUCTION}

\begin{table*}
\begin{center}
\begin{minipage}{150mm}
\caption{Program objects, observing conditions and results.}
\begin{tabular}{lcccccccccc}
Object name & $z$
&$M_B$
&$S_{1.4{\thinspace\rm GHz}}$
&$R_{8.4}$
&$t_{\rm int}$
&  $\tau_{225{\rm\thinspace GHz}}$ 
& $\tau_{850\mu{\rm m}}$ 
& $\tau_{450\mu{\rm m}}$ 
&$F_{850}$
&$F_{450}$
\\
&  
&
&(3$\sigma$)
&
&(mins) 
&(mean)
& 
& 
& (mJy) 
& (mJy) 
\\
\hline
BR B0019$-$1522 &4.52 &$-$27.6 &$<$0.5$^1$ &$<$2 &108 &0.08  & 0.21--0.26 & 1.1--2.4 
& 3.3$\pm$1.6 & 11$\pm$33\\
PSS J0134$+$3307 &4.52 &$-$28.2&$<$1.5$^2$ &$<$3 &108 &0.08 & 0.27--0.34 & 1.8--2.5 
& 0.4$\pm$1.5 & $-3\pm 54$\\
PSS J0248$+$1802 &4.43 &$-$27.9&$<$1.5$^2$  &$<$5 &36 & 0.10 & 0.23--0.46 & 2.4--3.2
&6.4$\pm$4.5 & $-433\pm264$\\
PSS J0747$+$4434 &4.42 &$-$28.2&$<$0.5$^3$  &$<$5 &54 &0.08 & 0.27--0.39 & 1.8--2.5 
& 0.0$\pm$2.2 & 72$\pm$105\\
BR B1600$+$0729 &4.35 &$-$28.1 &$<$0.5$^1$ &$<$2 &108 &0.09 & 0.28--0.33 & 1.7--2.0 
& 1.8$\pm$1.6 & $-7 \pm 55$\\
BR B2237$-$0607 &4.55 &$-$28.1 &$<$0.5$^1$ &$<$5 &144 &0.07 & 0.29--0.41 & 1.8--2.7 
& 5.0$\pm$1.1 & 19$\pm$53\\
\hline
\end{tabular}
\end{minipage}
\end{center}
Notes: Radio flux limits taken from (1) McMahon et al, in preparation;
(2) Condon et al, 1998; (3) Becker et al ,1995
\end{table*}

\noindent
Recent deep SCUBA surveys \cite{sib,hughes,barger,eales} have revealed
a significant population of discrete submillimetre sources, amounting
to $\sim$1000 deg$^{-2}$ above $\sim$3$\mJy$ at 850\mic. If
these objects lie at redshifts greater than 1, they have a far
infrared (FIR) luminosity $L_{\rm FIR}\sim 10^{12-13}\Lsun$, assuming
a dust temperature of $\sim$50K. Since
the large negative $K$-correction of the dust thermal spectrum means
that for $1\la z\la 10$, this luminosity is insensitive to source
redshift (\cf Arp220 in Figure~1). 
Prior to the SCUBA surveys, the only significant known
high redshift population with comparable bolometric luminosity were
the quasars. 
However, the characterisation of the FIR properties of
the high redshift quasars is rather meagre. Following the first
detections of  rest frame FIR emission from
z$>$3 radio quiet quasars \cite{mcmahon94}, our current
knowledge 
is still biased towards more
luminous objects with 850\mic\ fluxes in the
$\sim$10--50\mJy\ range \cite{isaak94,mcmahon94,omont96a}.

The onset of nuclear activity seems to be closely connected with
significant structure-formation events.  At high redshift,
black holes form in rare overdense regions and are fuelled by
their collapsing host objects \cite{hr}; at low redshifts,
interactions/mergers provoke inflow of gas to the galactic nuclei
\cite{bh}, fuelling central black holes.  These
processes would be expected to trigger a major starburst, though
the relative prominence of either component, starburst or AGN, would
depend on the precise dynamical details of the collapse or interaction.


In this paper, we report SCUBA 850\mic\ photometry of a statistically
complete sample of six of the most luminous, $z>4$, radio-quiet
quasars.  Our previous work at 800\mic\ with the JCMT (Isaak \etal
1994) and 1.25{\thinspace mm} with the IRAM 30{\thinspace m} (McMahon
\etal 1994, Omont \etal 1996a) and was only sensitive to relatively
high luminosity sources.  The aim of the current work is to carry out
more sensitive observations of a small sample of radio quiet quasars
in an effort to make an initial estimate of the contribution of AGN at
high redshift to the submillimetre source population.

Unless stated otherwise, we assume $H_0 = h_{50} \times 50 \km \s^{-1}
\Mpc^{-1}$ and $q_0 = 0.5$ throughout this paper.

\section{OBSERVATIONS}

The quasars for study were
selected from the APM samples of high redshift
quasars (Storrie-Lombardi \etal 1996, McMahon \etal in prep.)
and PSS survey (Kennefick \etal 1995). 
From extant radio data as
tabulated in Table~1, the quasars are radio quiet based on the
definition, Log $R_{8.4}$$<$1.0, where $R_{8.4}$=$L_{8.4}/L_B$,
is the ratio of radio and optical luminosities as defined by
Hooper et al(1996), where spectral indices of $-$0.5 and $-$0.7
have been used to extrapolate to the rest frame optical and 
radio fluxes.
Moreover, based on the observed
median spectral index of $-$0.7 found for radio loud quasars over
the wavelength range, the expected contamination at 850\mic\ is
$<$0.1mJy is all cases.
The most luminous $z>4$ objects were selected
in the HA range 15--08 as determined by the telescope scheduling,
providing effectively sparse sampling for  $M_B$$<$$-$27.5

Observations were carried out during three 8{\thinspace hr} observing
shifts over the period 1998 August 16 to 19, with the Submillimetre
Common-User Bolometer Array (SCUBA) \cite{holland99} on the
15{\thinspace m} James Clerk Maxwell Telescope. We used SCUBA in
standard point-source photometry mode at 450 and 850\mic. This
involves placing the target on the central bolometer of each array,
whilst `jiggling' the secondary mirror in a 9-point pattern with 2
arcsec offsets, integrating for 1 second at each. Superimposed on this
is a chop of the secondary mirror by 60 arcsec (in azimuth) at 7\Hz,
to remove sky emission; after the first 9-point jiggle, the telescope
is nodded such that the chop position lies on the opposite side of the
source. 
Uranus was used as the primary flux calibrator.
Skydips were taken regularly to determine the sky
opacity which varied considerably throughout the run as 
reported in Table~1.

Data reduction was performed with the Starlink {\tiny SURF} software,
and consisted primarily of experimenting with 
sky-removal techniques, using the mean and median of different rings
of bolometers as a background estimates.
The results were robust to variations in
the precise method, and we take the `mean-of-the-innermost-ring' as
our standard sky value.


\begin{figure}
\begin{center}
\psfig{figure=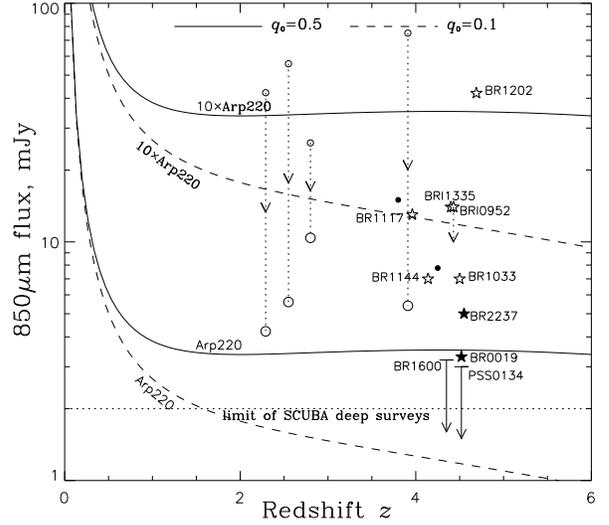,width=100.0mm,angle=0}
\end{center}
\caption{
  The current status of the high redshift ($z>$1) 850\mic\ Hubble
  diagram using the observations presented in Table~2.
Radio galaxies are shown as filled circles;
open circles show the submm detected lensed sources with
vertical lines connecting the observed flux with the unlensed
fluxes;
open stars show previously detected radio quiet $z>4$ quasars;
filled stars show our new observations. Upper limits are plotted
at the 2$\sigma$ limit with arrow length equal to the 
  1$\sigma$ uncertainty.
Also plotted is the expected flux from a thermal source like Arp220 as function of redshift for two cosmologies using the observed SED from 
Klaas et al.(1997).
%
}
\end{figure}

\begin{table*}
\begin{center}
\begin{minipage}{160mm}
\caption{Physical properties of representative 850\mic\ sources, assuming $T_{\rm d}=50\K$ and $\beta=1.5$}
\begin{tabular}{lccccccc}
Object name & &Redshift & $F_{850}$ & $M_{\rm{d}}$ & 
$L_{\rm FIR}\dagger$ & $SFR$ & $M_{B}\ddagger$\\
&&$z$&(mJy)&($h_{50}^{-2}10^8\Msun$)
&($h_{50}^{-2}10^{12}\Lsun$)
&($\alpha \times h_{50}^{-2}\Msunpyr$)&\\
\hline
BR0019$-$1522 &Quasar&4.52 & 3.3$\pm$1.6$^1$ & 1.5$\pm$0.7&5.0$\pm$2.4 & 500&$-$27.6\\ 
PSS0134$+$3307 &Quasar& 4.52 & $<3.0^1$ & $<1.4$& $<4.5$& $<450$&$-28.2$\\
BR1600$+$0729 &Quasar& 4.35 & $<3.2^1$ & $<1.5$& $<4.9$& $<490$&$-28.1$\\
BR2237$-$0607 &Quasar& 4.55 & 5.0$\pm$1.1$^1$ & 2.3$\pm$0.5& 7.5$\pm$1.6& 750&$-$28.1\\
&&&&&&\\
BRI0952$-$0115$^{l,a}$ &Quasar& 4.43 & 14$\pm$2$^2$& 6.5$\pm$0.9& 21.2$\pm$3.0&2120&$-$27.7 \\
BR1033$-$0327 &Quasar& 4.50 & 7$\pm$2$^2$ & 3.2$\pm$0.9 &10.5$\pm$3.0&1050&$-$27.6 \\
BR1117$-$1329 &Quasar& 3.96 & 13$\pm$1$^2$& 6.4$\pm$0.5&20.8$\pm$1.6&2080&$-$28.1\\
BR1144$-$0723 &Quasar& 4.14 & 7$\pm$2$^2$& 3.4$\pm$1.0&11.0$\pm$3.1&1100&$-$27.5 \\
BR1202$-$0725 &Quasar& 4.69 & 42$\pm$2$^2$& 19.0$\pm$0.9&62.0$\pm$3.0&6200&$-$28.5\\
BRI1335$-$0417 &Quasar& 4.40 & 14$\pm$1$^2$& 6.5$\pm$0.5&21.3$\pm$1.5&2130&$-$27.3\\
&&&&&&\\
8C1435$+$635 &Radio Galaxy& 4.25 &7.77$\pm$0.76$^3$&3.7$\pm$0.4&12.0$\pm$1.2&1200&$-$24.4\\ 
4C41.17 &Radio Galaxy& 3.80 & 17.4$\pm$3.1$^{4,m}$&7.5$\pm$1.3&24.5$\pm$4.4&2450&$-$25.0\\
APM08279$+$5255$^{l,b}$&Quasar&3.91&75$\pm$4$^5$&2.6$\pm$0.1&8.6$\pm$0.5&860&$-31.0$\\
SMM02399$-$0136$^{l,c}$&Seyfert 2&2.80&26$\pm$3$^6$&6.1$\pm$0.7&19.8$\pm$2.3&1980&$-$24.0\\
H1413$+$117$^{l,d}$&Quasar&2.55&66$\pm$7$^{7,m}$&3.4$\pm$0.4 &11.1$\pm$1.2&1110&$-28.5$\\
F10214$+$4724$^{l,e}$&Seyfert 2&2.29&50$\pm$5$^{8,m}$&0.9$\pm$0.1&2.9$\pm$0.3&290&$-23.1$\\
&&&&&&\\
Arp220&Starburst&0.0183&67000$^{9,n}$&1.7&5.5&550&$-21.3$\\
\hline
\end{tabular}
\end{minipage}

{\bf Notes:} 
$\dagger$ The uncertainty only reflects the flux uncertainty. 
$\ddagger$ Computed from observed optical continuum magnitudes 
assuming a spectral
index of $-$0.5, except in the cases of the radio galaxies,
where $K$-band magnitudes were used. 
$^l$Lensed objects. Magnification factors for the FIR continuum
are (a) uncertain; (b) 10( Green \& Rowan-Robinson, 1999; see
also Downes et al. 1995); 
(c) 2.5 (Ivison \etal, 1998b),
(d) 10 (Yun et al., 1997), (e) 30 (Eisenhardt et al. 1996). 
Derived physical quantities are corrected by these factors.
$^m$800\mic\ fluxes; equivalent 850\mic\ fluxes would be 15, 56 and 42\mJy, respectively. 
$^n${\it ISO} 170\mic\ flux ({\em N.B.}, at $z\sim4$, 850\mic\ corresponds 
to a rest wavelength of $\sim170$\mic).
\\
{\bf References:}
1. This work; 2. Buffey \etal (in prep.); 3. Ivison \etal (1998a); 
4. Dunlop \etal (1994); 5. Lewis \etal (1998); 6. Ivison \etal (1998b); 
7. Hughes \etal (1997); 8. Rowan-Robinson \etal (1993); 
9. Klaas \etal (1997).  
\end{center}
\end{table*}

\section{RESULTS}
At 850\mic, one quasar, 
BR2237$-$0607 at $z=4.55$, 
is detected with high significance (4.5$\sigma$),
while the quasar BR0019$-$1522 has a detection at 2$\sigma$. 
Two of the remaining objects have good
1$\sigma \approx 1.5$\mJy\ upper limits.
None of the objects are detected at
450\mic; at this wavelength, the observations are noisy, because weather
was on the whole unfavourable, so we disregard data in this band. 
Our observed fluxes are compared with those of other cosmological 
submillimetre sources in Figure~1.

We can derive dust masses and FIR luminosities 
from our observations as described in more detail in McMahon \etal (1994).
There is considerable uncertainty in these procedures (\eg Figure 2),
but
we can assume physical parameters derived for better-studied objects
and extrapolate from our 850\mic\ measurements. The mass of a dust
cloud optically-thin in the FIR is given by

\begin{equation}
M_{\rm{d}} = \frac{S_{\nu}(\nu_{\rm{obs}})D_{\rm{L}}^2}{\kappa_d B_{\nu}(\nu_{\rm{rest}},T_{\rm{d}})(1+z)},
\end{equation}

\noindent where $D_{\rm{L}}$ is the luminosity distance
and $\kappa_d$ is the absorption coefficient: 

%



\begin{equation}
\kappa_{\rm d} = \kappa_{850} \times 
\left(\frac{\nu_{\rm rest}}{353{\rm\thinspace GHz}}\right)^{\beta}.
\end{equation}

\noindent We assume a frequency dependence of absorption coefficient 
of $\beta = 1.5$, as shown to be a good fit for
both low redshift luminous infrared galaxies (Carico
\etal, 1992) and luminous high redshift quasars
(Benford \etal, in prep.) and $\kappa_{850} = 0.11\m^2 \kg^{-1}$, based on the
widely-used Hildebrand (1983) normalisation at 125\mic($\kappa$=1.875$\rm m^2/kg$), 
which we have 
extrapolated
to longer rest wavelengths assuming $\beta$=1.5.


The above assumes that the rest-frame FIR continuum is
thermal emission from dust which is optically thin to FIR photons,
which is justified since spectral indices in the Rayleigh--Jeans
region for typical $z>4$ quasars are $\ga 3$ (Buffey \etal in prep.,
Isaak \etal, 1994).
It also assumes that the dust is isothermal: a single temperature of
50\K\ is a typical value derived from fits to well-covered FIR SEDs.
If the dust were not isothermal, fits to the currently-available data
would be biased towards the higher temperatures,
causing us to underestimate the true dust mass. $ISO$
Photometry of nearby ULIRGs \cite{klaas} suggests that
two-component fits are more appropriate, low-temperature dust at
$\sim$50\K, presumed to be heated by starburst, and high-temperature dust
at $\sim$150\K, possibly heated by an AGN; whereas observations of
quasars \cite{haas} suggest much higher ($150\K \la T_{\rm d} \la 600\K$)
 temperatures for
this component. Since the
existing $z>4$ data do not constrain two-component fits, we assume
that the 850\mic\ flux is dominated by the low-temperature component.

The FIR luminosity is obtained by integrating under a 
single modified blackbody curve. 
Given, $T_{\rm d}$=50K and $\beta$=1.5, $L_{\rm FIR}$ is 
directly proportional to $M_{\rm d}$: 

\begin{equation}
L_{\rm FIR} \approx 3.3 \times \left( \frac{M_{\rm d}}{10^8\Msun} \right) \times 10^{12}\Lsun.
\end{equation}
%
%
Assuming that the FIR flux is due to dust heated by a starburst,
the star formation rate is given by:
\begin{equation}
SFR = \alpha \times 10^{-10}\frac{L_{\rm FIR}}{\Lsun}\Msunpyr,
\end{equation}
where the value of $\alpha$ depends
upon the stellar initial mass function (IMF) adopted and the 
timescale for the burst.
For example, Scoville \& Young
(1983) deduce $\alpha=0.77$ by considering the total energy radiated by
main sequence O, B and A stars ($M > 1.6\Msun$), 
over $\sim 10^9\yr$; whereas Thronson \& Telesco (1986) obtain
$\alpha=2.1$, assuming a Salpeter IMF.
Allowing for the low mass (1.6$\Msun > M > 0.1\Msun$) stars, 
$\alpha$ rises by a further factor $\sim$3. 


Table 2 shows results derived for a sample of cosmological
submillimetre sources, 
in a self-consistent manner,
to enable comparison of observed and derived properties.
The uncertainties inherent in our assumptions  are illustrated
in Figure 2.
Note that the uncertainties
in FIR luminosities reflect solely the uncertainty in the
flux, and do not include the uncertainty due to
the assumed single temperature.
In Figure 3 are
plotted the inferred FIR luminosities against optical luminosity
for the 
z$>$4 quasars.
Since the quasar BR1202$-$0725 has
been resolved into two millimetre and optical sources \cite{omont96b,hme} 
it is plotted as
two discrete points, since its extreme luminosity might be explained if it 
represents the merger of two gas-rich galaxies or 
of an AGN and a gas-rich galaxy.

\begin{figure}
\psfig{figure=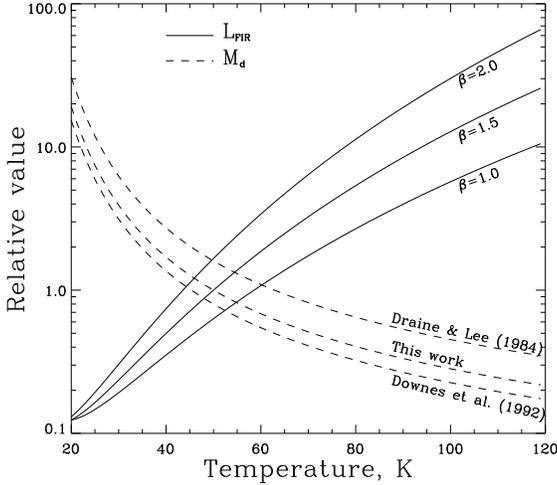,width=90mm,angle=0.0}
\caption{
  Variation of derived dust mass and FIR luminosity with dust
  temperature, relative to values with $T=50\K$ and $\beta=1.5$, for
  various values of $\beta$ and absorption coefficient, calculated from a 850\mic\ flux from a dust cloud at $z=4.5$.}
\end{figure}

\begin{figure}
  \psfig{figure=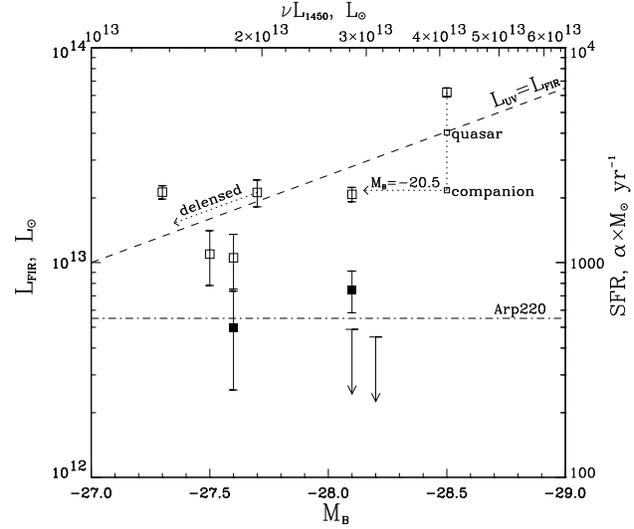,width=90mm,angle=0.0}
\caption{
FIR luminosity and inferred star formation rate
against absolute $B$-band magnitude for the z$>$4 quasars
tabulated in Table~2.
Filled squares and upper limits: this work. 
Open squares: 
previously studied quasars.
The dot-dashed horizontal line shows the FIR luminosity of Arp220. The dashed
line shows the locus of $L_{\rm FIR} = L_{\rm UV}$. 
%
} 
\end{figure}

\section{DISCUSSION}

From the results presented in Figure~3, there is no obvious
correlation between $L_{\rm FIR}$ and $L_{\rm UV}$.  Indeed, over a
range of a factor of 1.5 in UV luminosity, the scatter in $L_{\rm FIR}$ 
is $\sim10$.  The absence of a strong correlation
between $L_{FIR}$ and $L_{UV}$ is supported by the results tabulated
in Table~2, with the obvious caveat that this compilation, 
is rather
heterogeneous and consists primarily of detected objects.
This conclusion has the
implication that the high redshift AGN population may have a FIR
luminosity function that is independent of UV luminosity.

In earlier mm/submm work \cite{isaak94,omont96a}, only a small
fraction ($\sim20$ percent) of the observed sources were detectable,
and the observational limits were at quite high FIR luminosities.
These new observations indicate that the typical luminous quasar has a
FIR luminosity less than or equal to that of Arp220.

We can
investigate the AGN contribution to the deep SCUBA source counts
using a simple model
where all AGN have 
a fiducial FIR luminosity of $\sim5\times10^{12}\Lsun$ with an Arp220-like
spectrum in the submm range.
We assign this FIR luminosity to each quasar and
adopt a standard Boyle, Shanks \& Peterson (1988) quasar
luminosity function, extrapolating to high redshifts by assuming no
evolution between $z=2$ and $z=3$, and a density fall-off by a factor of
2 per unit redshift thereafter.  Integrating the 
luminosity
function down to $M_B=-$23.0, we find that there are $\sim$140
quasars deg$^{-2}$. 
This absolute magnitude limit is
primarily based on historical convention from ground based image
quality, and if we relax this
limit and extrapolate down to $M_B=-$21.0, the source
density reaches $\sim$400 deg$^{-2}$.
Finally, if one
allows for the 
observed ratio of Seyferts II to Seyferts I at low redshift 
(2.3$\pm$0.7; Huchra \& Burg, 1992), even neglecting any
contribution from totally optically obscured AGN, it is
feasable that
that a major part (15--100 per cent) 
of the SCUBA source population 
could possess AGN.


Further evidence for such high space densities of AGN
comes from the deepest X-ray surveys which reach source
densities of $\sim$1000 deg$^{-2}$ (Hasinger \etal 1998).
Whilst the
nature of the faintest sources is still controversial, an AGN origin
for the energy source is widely considered as the most 
plausible explanation (\eg Hasinger \etal 1998).
Furthermore, estimates of the local space density of 
massive dark objects within galaxies \cite{magorrian} 
exceed the density of the optically-selected quasars, of which 
these objects are presumed to be the dull remnants  \cite{hnr,tbg99}:
there may
exist a significant population of AGN which are either obscured
by dust or accrete in low-efficiency modes.

Lower limits on the contribution of AGN to the submm source
counts comes from studies of local ultraluminous infrared galaxies.
These limit the fraction of ULIRGs in which AGN activity
is the primary energy source to 20--30 percent, 
although these observations do not 
exclude the possibility that the remaining starburst-dominated
cases contain AGN as well\cite{genzel98}.


An alternative estimate of the contribution of broad
line AGN to the submm source counts can be derived by assuming that 
$L_{FIR}$ is linearly correlated with $L_{UV}$ as shown in Figure 3. 
Above 3mJy at 850\mic,
the source density ranges between 4 and 9 deg$^{-2}$ for a thermal 
spectrum with temperature of 70K and 30K respectively. This is
essentially because only AGN with $M_B$$<$$-$25 would have detectable
flux at 850\mic.  It is thus important that observations of quasars
with $M_B$$<$$-$25 are carried out at 850\mic\ to distinguish 
between the range of models.

If AGN are present in a significant fraction of the SCUBA 
sources, this has important implications for studies which use the
SCUBA source counts to infer star formation rates from the
from a 850\mic\ flux. 
The tacit assumption in Equation 3 is that this locally-derived
relation between $L_{\rm FIR}$ and the star formation rate, based
on galaxies such as M51 in which $L_{\rm FIR} \sim 10^{10}\Lsun$, 
can be applied within the extreme environments of high redshift
AGN and massive starbursts, in which FIR luminosities are 100--1000
times larger. For example, the intense UV flux from any starburst
or AGN could suppress the formation of low mass stars which dominate
the total mass. Thus, even if the dust is not heated by a more 
energy-efficient AGN, 
the AGN's presence
may effect the underlying IMF which is used to deduce the
total star formation rate.

\section{CONCLUSIONS}
We have observed a small but statistically-complete 
sample of luminous high redshift ($z>4$) quasars with SCUBA at 850 and
450\mic, and have found that the typical FIR luminosity of these
quasars is less than or comparable with that of the  archetypal local
ultraluminous IRAS galaxy Arp220. 
Using a simple model for the contribution of AGN to the submm source 
population, we find that $\ga15$ per cent of the sources in the deep 
SCUBA surveys
at the $S_{850}$$\sim$3-4mJy level, 
may contain active nuclei.
Hence, both the contribution of AGN to the FIR luminosity, 
and its effect upon the underlying IMF (should starburst be the
dominant power source), need to be considered.

Further submm studies of lower redshift and lower luminosity quasars
are required in order to determine directly the far infrared 
luminosity function
for quasars. In addition sensitive  X-ray observations with {\it AXAF} and 
{\it XMM} may 
determine the fraction of the SCUBA sources which harbour AGN.



\section*{ACKNOWLEDGMENTS}
We thank Kate Isaak and the referee for comments which
helped us clarify the contents of this paper.
RGM thanks the Royal Society for support, and RSP acknowledges receipt of
a PPARC studentship. IS acknowledges support under the European Commission, 
TMR Programme, Research Network Contract ERBFMRXCT96-0034 ``CERES''.
The JCMT is operated by the Joint Astronomy
Center on behalf of the UK Particle Physics and Astronomy Research
Council, the Netherlands Organisation for Scientific Research and the
Canadian National Research Council.

\end{document}